\documentclass[3p,times,twocolumn]{elsarticle}
\pdfoutput=1
\usepackage{ecrc}
\usepackage{epsfig}
\usepackage{graphicx}
\usepackage[bookmarks=false]{hyperref}

\volume{00}

\firstpage{1}

\journalname{Nuclear Physics B Proceedings Supplement}

\runauth{}

\jid{nuphbp}

\jnltitlelogo{Nuclear Physics B Proceedings Supplement}

\usepackage{amssymb}

\usepackage[figuresright]{rotating}

\begin{document}

\def\a{\alpha}
\def\b{\beta}
\def\c{\chi}
\def\d{\delta}
\def\e{\epsilon}
\def\f{\phi}
\def\g{\gamma}
\def\h{\eta}
\def\i{\iota}
\def\j{\psi}
\def\k{\kappa}
\def\l{\lambda}
\def\m{\mu}
\def\n{\nu}
\def\o{\omega}
\def\p{\pi}
\def\q{\theta}
\def\r{\rho}
\def\s{\sigma}
\def\t{\tau}
\def\u{\upsilon}
\def\x{\xi}
\def\z{\zeta}
\def\D{\Delta}
\def\F{\Phi}
\def\G{\Gamma}
\def\J{\Psi}
\def\L{\Lambda}
\def\O{\Omega}
\def\P{\Pi}
\def\Q{\Theta}
\def\S{\Sigma}
\def\U{\Upsilon}
\def\X{\Xi}

\def\ve{\varepsilon}
\def\vf{\varphi}
\def\vr{\varrho}
\def\vs{\varsigma}
\def\vq{\vartheta}
\def\dg{\dagger}                                     
\def\ddg{\ddagger}                                   
\def\wt#1{\widetilde{#1}}                    
\def\mt{\widetilde{m}_1}
\def\mti{\widetilde{m}_i}
\def\rt{\widetilde{r}_1}
\def\mtt{\widetilde{m}_2}
\def\mttt{\widetilde{m}_3}
\def\rtt{\widetilde{r}_2}
\def\mb{\overline{m}}
\def\VEV#1{\left\langle #1\right\rangle}        
\def\be{\begin{equation}}
\def\ee{\end{equation}}
\def\ds{\displaystyle}
\def\ra{\rightarrow}

\def\bea{\begin{eqnarray}}
\def\eea{\end{eqnarray}}
\def\NO{\nonumber}
\def\Bar#1{\overline{#1}}

\begin{frontmatter}




\dochead{}

\title{Developments in Leptogenesis
 \tnotetext[label1]{Talk at Neutrino 2010. Dedicated to the memory of Alexey Anisimov (http://www2.physik.uni-bielefeld.de/714.html).}
}


\author{Pasquale Di Bari}

\address{School of Physics and Astronomy, University of Southampton, Southampton, SO17 1BJ, U.K., \\
Department of Physics and Astronomy, University of Sussex, Brighton, BN1 9QH, U.K.}

\begin{abstract}
Latest developments in leptogenesis are reviewed
with a particular emphasis on the proposals to test leptogenesis.
We discuss in particular the important role played by light and heavy flavour effects
in the determination of the final asymmetry and the attractive features of the $N_2$
dominated scenario.
\end{abstract}


\end{frontmatter}



\section{The double side of leptogenesis}

Leptogenesis \cite{fy} realizes a highly non trivial link
between two completely independent experimental observations:  a global
property of the Universe, the absence of primordial anti-matter
in the observable Universe and the observation that neutrinos mix and (therefore) have masses.
In this way leptogenesis has a naturally built-in double sided nature.
On one hand it describes a very early stage in the history of the Universe characterized by temperatures
($T_{\rm Lep}\gtrsim 100\,{\rm GeV}$)
much higher than those probed by Big Bang Nucleosynthesis ($T_{BBN} \sim 1\,{\rm MeV}$). On the other hand
leptogenesis  complements low energy neutrino experiments providing a completely independent phenomenological
tool for testing the high energy parameters in the seesaw mechanism \cite{seesaw}.
In these proceedings we will mainly focus on this second side of leptogenesis,
where the early Universe history is basically exploited as a neutrino physics experiment.

\section{Vanilla leptogenesis and beyond}

\subsection{Vanilla leptogenesis}

Leptogenesis is a (cosmo)logical consequence of the the seesaw mechanism
that elegantly explains not only
why neutrinos mix and have masses but also why they are so much lighter than all the other
massive fermions. In a minimal  type I
seesaw mechanism  right-handed neutrinos with neutrino Yukawa coupling  $h$
and a right-right Majorana mass term are added to the Standard Model Lagrangian,
\bea\label{lagrangian}
\mathcal{L} & = & \mathcal{L}_{\rm SM} +i \overline{N_{R i}}\g_{\m}\partial^{\m} N_{Ri} -
h_{\a i} \overline{\ell_{L\a}} N_{R i} \tilde{\F} - \\
& & {1\over 2}\,M_i \overline{N_{R i}^c}N_{R i} +h.c. \nonumber
\eea
$(i=1,2,3,\quad \a=e,\m,\t)$.
For definiteness we consider the case of three RH neutrinos species.
This is also the most attractive option with
one RH neutrino for each family, as nicely predicted by $SO(10)$ grand unified
models. Notice however that all current data from low energy neutrino experiments
are consistent  with a more minimal two RH neutrino model.

After spontaneous symmetry breaking, a Dirac mass term $m_D=v\,h$
is generated by the  Higgs vev $v$. In the seesaw limit, $M\gg m_D$, the spectrum
of neutrino masses splits into a light set given by the eigenvalues $m_1<m_2<m_3$
of the neutrino mass matrix
\be\label{seesaw}
m_{\nu} = - m_D\,{1\over M}\,m_D^T \,
\ee
and into a heavy set $M_1 <  M_2 < M_3$ coinciding with very good
approximation with the eigenvalues of the Majorana mass matrix.
The symmetric neutrino mass matrix $m_{\nu}$ is diagonalized by a unitary matrix $U$,
$D_m \equiv {\rm diag}(m_1,m_2,m_3) = -U^{\dagger}\,m_{\nu}\,U^{\star}$ that
in a basis where the charged lepton mass matrix is diagonal
coincides with the leptonic mixing matrix $U_{PMNS}$.

In this way the lightness of ordinary neutrinos is explained just
as an algebraic by-product. If the largest eigenvalue in the Dirac neutrino mass matrix
is assumed to be of the order of the electroweak scale, as for the other massive fermions,
then the atmospheric neutrino mass scale $m_{\rm atm}\equiv \sqrt{m_3^2-m^2_1}\simeq 0.05\,{\rm eV}$ can be
naturally reproduced for $M_3 \sim 10^{14-15}\,{\rm GeV}$, very close to the grand-unified scale.

In general, the decays of the right-handed neutrinos  violate $C\!P$
so that the decay rate $\Gamma_i$ for $N_i \rightarrow l_i + \phi^{\dagger}$
can be different from the decay rate $\bar{\Gamma}_i$ for $N_i \rightarrow \bar{l}_i + \phi$.
In this way each $N_i$-decay will produce, on average, a B-L number given by
the total $C\!P$ asymmetry defined as
$\ve_i \equiv -{(\Gamma_i - \bar{\Gamma}_i)/(\Gamma_i + \bar{\Gamma}_i)}$ .
If the $N_i$'s decay at temperatures $T\gtrsim 100\,{\rm GeV}$,
then non-perturbative ($B-L$ conserving) sphaleron processes are in equilibrium ($\Gamma_{\rm sph}\gtrsim H$)
so that lepton and baryon numbers are not separately conserved and $N_B\simeq N_{B-L}/3$.
The final baryon-to-photon number ratio can then be calculated from
the final $B-L$ asymmetry as $\eta_B \simeq 0.01 \, N_{B-L}^{\rm f}$,
where we indicate with $N_X$ the value of any quantity $X$  in a portion of comoving volume
that contains one RH neutrino in ultra-relativistic thermal equilibrium.
The result has to be compared with the measured value from CMB anisotropies observations \cite{WMAP7}
\be\label{etaBobs}
\eta_B^{\rm CMB} = (6.2 \pm 0.15)\times 10^{-10} \, .
\ee
In a minimal version of leptogenesis a type I seesaw mechanism is assumed together
with a thermal production of the RH neutrinos from the scatterings of particles in the thermal bath
(thermal leptogenesis). In this way a non negligible RH neutrino $N_i$-abundance
requires $T > {\cal O}(M_i)$.

The final asymmetry has been traditionally calculated in a very simple
way neglecting both the flavour composition of the lepton quantum states
produced by $N_i$-decays (light flavour effects) and the production of the
asymmetry from the heavier RH neutrino decays (heavy flavour effects). In this
oversimplified picture, that we  call {\em vanilla leptogenesis},
the final asymmetry is then simply given by $N_{B-L}^{\rm f}\simeq \ve_1\,\kappa^{\rm f}(K_1)$,
where $K_1 \equiv (\G + \bar{\G})/H(T=M_1)$ is the lightest RH neutrino
decay parameter and $\k^{\rm f}(K_1)$ is the final efficiency factor
giving approximately the number of
$N_1$'s decaying out-of-equilibrium.

Barring fine tuned cancelations among the
terms giving the RH neutrino masses in the see-saw formula,
the total $C\!P$ asymmetry is upper bounded by \cite{di},
\be\label{CPbound}
\ve_1 \leq \ve_1^{\rm max} \simeq 10^{-6}\,{M_1\over 10^{10}\,{\rm GeV}}\,{m_{\rm atm}\over m_1+m_3} ,
\ee
and, imposing $\eta_B^{\rm max}\simeq 0.01\,\ve_1^{\rm max}\,\k_1^{\rm f} > \eta_{B}^{CMB}$,
one obtains, in the plane $(m_1,M_1)$, the allowed region   shown in Fig.~1.
\begin{figure}
\vspace*{-1mm}
\begin{center}
     \psfig{file=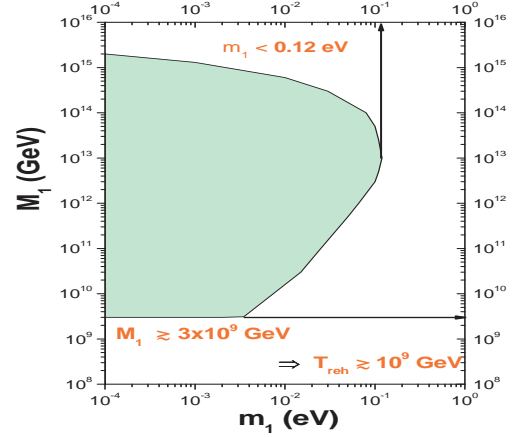,height=65mm,width=75mm}
     \vspace*{-3mm}
     \caption{Neutrino mass bounds in the vanilla scenario.}
\end{center}
\vspace*{-7mm}
\end{figure}
One can notice the existence of an upper bound on the light
neutrino masses $m_1\lesssim 0.12\,{\rm eV}$, incompatible with quasi-degenerate
neutrino mass models, and a lower bound on
$M_1\gtrsim 2\times 10^9\,{\rm GeV}$ \cite{di} implying a lower bound on the
reheat temperature $T_{\rm RH}\gtrsim 10^9\,{\rm GeV}$ \cite{pedestrians}.
These bounds are valid under the following set of assumptions and approximations \cite{bounds}:
i) the  flavour composition of the leptons in the final states is neglected;
ii) the heavy RH neutrino mass spectrum is assumed to be strongly hierarchical
     with $M_2\gtrsim 10\,M_1$;
iii) there is no interference between the heaviest RH neutrino and the
       next-to-lightest RH neutrino, i.e. $(m^{\dagger}_D\,m_D)_{23}=0$.
The last two conditions guarantee
that $\ve_{2,3}^{\rm max}\,\kappa(K_{2,3})\ll \ve_{1}^{\rm max}\,\kappa(K_1)$.
In particular, the last condition is always verified for $M_3\gg 10^{14}\,{\rm GeV}$,
when an effective two RH neutrino model is recovered.

An important feature of vanilla leptogenesis is that the final asymmetry does not
directly depend on the parameters of the leptonic mixing matrix and therefore
one cannot establish any kind of direct connection. In particular a discovery
of CP violation in neutrino mixing would not be a smoking gun for leptogenesis
but on the other hand a non discovery would not rule out leptogenesis.
However, within more restricted scenarios, where for example some conditions on
the neutrino Dirac mass matrix are imposed, links can emerge. We will discuss
in detail the case of $SO(10)$-inspired models.

Many different directions have been explored in order to go beyond
the assumptions and the approximations of the vanilla leptogenesis scenario,
often with the objective of finding ways to evade the bounds shown in Fig.~1.
 Let us briefly discuss the main results.

\subsection{Beyond a hierarchical RH mass spectrum}

If $(M_2-M_1)/M_1 \equiv \delta_2 \ll 1$, the $C\!P$ asymmetries get resonantly
enhanced as $\ve_{1,2}\propto 1/\d_2$. If, more stringently, $\d_2\lesssim 10^{-2}$, then
$\eta_B \propto 1/\delta_2$ and the degenerate limit is obtained.
In this limit the lower bounds on $M_1$ and on $T_{\rm RH}$
get relaxed $\propto \delta_2$ and at the resonance they completely disappear \cite{beyondHR}.
However, there are not many models able to justify in a reasonable way such a degenerate limit.
 Examples are provided by radiative leptogenesis and by models with extra-dimensions
where all RH neutrinos masses squeeze together to a common  TeV scale \cite{DLmodels}.

\subsection{Non minimal leptogenesis}

Other proposals to relax the lower bounds on $M_1$ and on $T_{RH}$
rely on extensions  beyond minimal leptogenesis. For example on the
 addition of a right-right Majorana mass term yielding a  type II seesaw mechanism \cite{typeII}
or on a non thermal production of the RH neutrinos whose decays produce the asymmetry \cite{nonth}.
However, these non minimal models spoil somehow a remarkable coincidence between the measured values
of the atmospheric and solar neutrino mass scales and the possibility to have successful leptogenesis
even independently of the initial conditions \cite{pedestrians,bounds}.
Non minimal models have been also extensively explored in order to get a
low scale leptogenesis testable at colliders \cite{mohapatratalk}.

\subsection{Improved kinetic description}

Within vanilla leptogenesis the asymmetry is calculated solving simple rate equations,
classical Boltzmann equations integrated over the RH neutrino momenta.
Different kinds of extensions have been studied, for example accounting for a full momentum dependence
\cite{momentum}, for quantum kinetic effects \cite{quantum} or for thermal effects \cite{thermal}.
All these analyses find  significant changes in the weak wash-out regime but within
$\sim 50\%$ in the strong wash-out regime. This result has
quite a straightforward general explanation. In the strong wash-out regime the final asymmetry
is produced by the decays of RH neutrinos in a non relativistic regime \cite{pedestrians}
when a simple classical momentum independent kinetic description provides quite a good approximation.
It should therefore be bourne in mind that the use of a simple kinetic description in leptogenesis
is not just a simplistic approach but is justified in terms of the
neutrino oscillations experimental results on the neutrino masses that
support a strong wash-out regime.

\section{Flavour effects}

In the last years, flavour effects proved to be the most relevant
modification  of the vanilla scenario and for this reason we
discuss them in a separate section.
There are two kinds of flavour effects that are neglected in the vanilla scenario: heavy
flavour effects \cite{geometry}, how heavier RH neutrinos influence the final asymmetry,
and light flavour effects \cite{flavoreffects}, how the
flavour composition of the leptons quantum states produced in the RH neutrino decays
influence the final asymmetry.
We first discuss the two effects separately and then we show
how their interplay has a very interesting application \cite{vives}.

\subsection{Light flavour effects}

Let us first start by continuing to assume that the final asymmetry is
dominantly produced by the decays
of the lightest RH neutrinos, neglecting the contribution from the
decays of the heavier RH neutrinos.
If $M_1\gtrsim 5\times 10^{11}\,{\rm GeV}$, the flavour composition
of the quantum states of the leptons produced in $N_1$ decays
has no influence on the final asymmetry and the unflavoured regime holds.
This is because the lepton quantum states evolve coherently  between the production of a lepton from an $N_1$-decay
and a subsequent inverse decay with an Higgs boson. In this way
the lepton flavour composition does not play any role.

However, if $5\times 10^{11}\,{\rm GeV}\gtrsim M_1 \gtrsim 10^{9}\,{\rm GeV}$, then
between one decay and the subsequent inverse decay, the produced lepton quantum states, on average,
interact with tauons in a way that the coherent evolution breaks down. Therefore, at
the inverse decays, the leptons quantum states are an incoherent mixture of a tauon component and
of a (still coherent) superposition of an electron and of a muon component that we will
indicate with $\gamma$.
The fraction of asymmetry stored in each flavour component is not proportional in general
to the branching ratio of that component. This implies that the dynamics of the two
flavour asymmetries, the tauon and the $\gamma$ asymmetries, are different and have to be separately
calculated. In this way  the resulting final asymmetry can considerably differ
from the result in the unflavoured regime.
If $M_1\lesssim 10^{9}\,{\rm GeV}$, then even the coherence of the $\gamma$
component is broken by the muon interactions between decays and inverse decays
and a full three flavour regime applies. In the intermediate regimes
a density matrix formalism is necessary to describe the transition \cite{flavoreffects,densitymatrix}.

There are three kinds of major modifications induced by flavour effects.
First, the wash-out can be  considerably lower than in the unflavoured regime \cite{flavoreffects}.
Second, the low energy phases affect directly the final asymmetry since they
contribute to a second source of $C\!P$ violation in the flavoured $C\!P$ asymmetries
\cite{ibarra,flavorlep,ppr}. As a consequence the same source of $C\!P$
violation that could take place in neutrino oscillations, could be also responsible for the observed
matter-antimatter asymmetry of the Universe, though under quite stringent
conditions on the RH neutrino mass spectrum \cite{diraclep}.
A third modification is that the flavored $C\!P$ asymmetries contain
extra-terms that evade the upper bound eq.~(\ref{CPbound}) if some
mild cancelations in the seesaw formula among the light neutrino mass terms and
just a mild RH neutrino mass hierarchy ($M_2/M_1 \lesssim 10$) are allowed. In this way the lower bound on the
reheat temperature can be relaxed by about one order of magnitude, down to $10^8\,{\rm GeV}$
\cite{bounds} (see Fig.~2).
\begin{figure}
\vspace*{-15mm}
\begin{center}
     \psfig{file=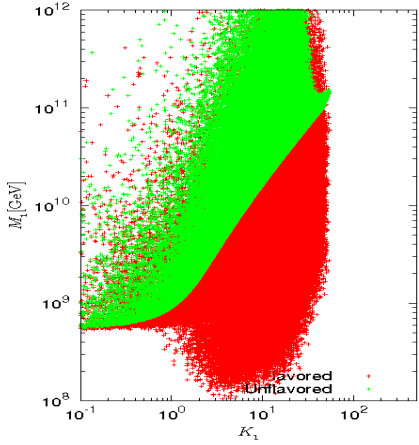,height=95mm,width=105mm}
     \vspace*{-3mm}
     \caption{Relaxation of the lower bound on $M_1$ thanks
     to additional unbounded flavoured $C\!P$ violating terms.}
\end{center}
\vspace*{-7mm}
\end{figure}

\subsection{Heavy flavour effects}

In the vanilla scenario the contribution to the final asymmetry from the
heavier RH neutrinos is negligible for two reasons: the $C\!P$ asymmetries of $N_2$ and $N_3$ are
suppressed in the hierarchical limit with respect to $\ve_1^{\rm max}$ and
even assuming that a sizeable asymmetry is produced around $T \sim M_{2,3}$,
this is later on washed out by the lightest RH neutrino
inverse processes. However, it has been realized that the assumptions
for the validity of the vanilla scenario are quite restrictive and
there are a few reasons why heavy flavour effects have to be taken into account
in general.

First, in the quasi-degenerate limit when $\d_{2,3} \ll 1$,
the $C\!P$ asymmetries are not suppressed and the wash-out from the lightest RH neutrinos
is only partial \cite{beyondHR}.
Second, even assuming a strong RH neutrino mass hierarchy, there is always
a choice of the parameters such that $N_1$ decouples and its wash-out vanishes.
For the same choice of the parameters, the $N_2$ total $C\!P$ asymmetry is unsuppressed
if $M_3\lesssim 10^{15}\,{\rm GeV}$ . In this
case a $N_2$-dominated scenario is realized \cite{geometry}.
Notice that the existence of a third heaviest RH neutrino species is crucial.
Third, even assuming a strong mass hierarchy, a coupled $N_1$ and $M_1\gtrsim 10^{12}\,{\rm GeV}$,
the asymmetry produced by the heavier RH neutrino decays, in particular by the $N_2$'s decays, with unsuppressed
total $C\!P$ asymmetry can be sizeable and in general is not completely washed-out by the lightest RH neutrino
processes. This is because there is in general a component
that escapes the $N_1$ wash-out \cite{bcst,nardinir}. Notice that for a mild mass hierarchy,
$\d_3 \lesssim 10$,
even the asymmetry produced by the $N_3$'s decays can be sizeable and circumvent
 the $N_1$ and $N_2$ wash-out.

\subsection{Flavoured $N_2$ dominated scenario}

There is an another interesting scenario where the asymmetry from the $N_2$ decays
dominates the final asymmetry. This scenario relies on the interplay between
light and heavy flavour effects \cite{vives}.
Even assuming a strong mass hierarchy, a coupled $N_1$ and $M_1\lesssim 10^{12}\,{\rm GeV}$,
the $N_1$ wash-out  can be circumvented. Suppose for example that the
lightest RH neutrino wash-out occurs in the three-flavour regime ($M_1 \ll 10^{9}\,{\rm GeV}$).
In this case the asymmetry produced by the
heavier RH neutrinos, at the $N_1$ wash-out, distributes into an incoherent mixture of
light flavour quantum eigenstates. It turns out that
the $N_1$ wash-out in one of the three flavours is negligible in quite a wide region of the parameter space.
In this way, accounting for flavour effects, the region of applicability  of the
$N_2$-dominated scenario  enlarges considerably, since it is not
necessary that $N_1$ fully decouples but it is sufficient that it decouples
just in one particular light flavor. Recently, it has been realized that,
accounting for the Higgs and for the quark asymmetries, the dynamics of the flavour asymmetries
couple and the lightest RH neutrino wash-out in a particular flavour can be circumvented even when $N_1$ is strongly
coupled in that flavour \cite{flcoupling}.
Another interesting effect arising in the $N_2$-dominated scenario is phantom leptogenesis.
This is a pure quantum-mechanical effect that for example
allows  parts of the electron and of the muon asymmetries, the phantom terms, to escape completely
the wash-out at the production when $T\sim M_2 \gg 10^{9}\,{\rm GeV}$.

\section{Testing new physics with leptogenesis}

The seesaw mechanism extends the Standard Model introducing eighteen new parameters
when three RH neutrinos are considered. On the other hand, low energy
neutrino experiments can only potentially test nine parameters in the
neutrino mass matrix $m_{\nu}$. Nine high energy parameters, those characterizing the properties
of the three RH neutrinos (three life times, three masses and three total $C\!P$ asymmetries)
and encoded in the orthogonal matrix $R$ \cite{casas},
are not tested by low energy neutrino experiments.
Quite interestingly, leptogenesis gives an additional constraint on a combination
of both low energy neutrino parameters and high energy neutrino parameters,
$\eta_B(m_{\nu},R)=\eta_{B}^{CMB}$. However,
just one additional constraint does not seem to be still sufficient to over-constraint the parameter
space leading to testable predictions. Despite this, as we have seen, in the vanilla leptogenesis scenario
there is an upper bound on the neutrino masses. The reason is that in this case
$\eta_B$ does not depend on the 6 parameters related to the properties of the two heavier RH neutrinos and
therefore the asymmetry depends on a reduced number of high energy parameters. At the
same time, the final asymmetry is strongly suppressed by the absolute neutrino mass scale when this is
larger than the atmospheric neutrino mass scale. This is why the leptogenesis
bound yields an upper bound on the neutrino masses.

When flavour effects are considered, the vanilla leptogenesis
scenario holds only under very special conditions. More generally
the parameters in the leptonic mixing matrix also
directly affect the final asymmetry and, accounting for flavour effects,
one could hope to derive definite predictions on the leptonic mixing matrix .
However, when flavour effects are taken into account,
the 6 parameters associated to the two heavier RH neutrinos contribute in general to  the  final
asymmetry at the expenses of predictability.
For this reason, in a generic scenario with three RH neutrinos, it is not possible
to derive any prediction on low energy neutrino parameters.

In order to gain predictive power, two possibilities have been largely explored in the last years.
In a first case one considers non minimal scenarios giving rise to
additional phenomenological constraints.
We have already mentioned how with a non minimal seesaw mechanism it is possible to lower
the leptogenesis scale and have signatures at colliders. It has also been noticed that
in supersymmetric models one can enhance the branching ratios of lepton flavour violating processes
or electric dipole moments and in this way the existing experimental bounds
further constrain the seesaw parameter space \cite{lfvedm}.

A second possibility is to search again, as within vanilla leptogenesis, for a reasonable
scenario where the final asymmetry depends on a reduced number of free parameters in a way that the
parameter space gets over-constrained by the leptogenesis bound. Let us briefly discuss some
of the ideas that have been proposed.

\subsection{Two RH neutrino model}

A  well motivated scenario that attracted great attention is a two
RH neutrino scenario \cite{2RHN}, where the third RH neutrino is either absent or
effectively decoupled. This necessarily happens when $M_3\gg 10^{14}\,{\rm GeV}$, implying that the
lightest LH neutrino mass $m_1$ has to vanish. It can be shown that the number of parameters
gets reduced from 18 to 11. It has been shown that in this case
inverted hierarchical models with
$\sin\theta_{13}\cos\delta \gtrsim -0.15$ are viable only if there
is $C\!P$ violation from Majorana  phases \cite{mp}.
However this prediction would be very difficult to test and in any case
would be quite unlikely to provide a smoking gun.

\subsection{$SO(10)$ inspired models}

The only way to gain a strong predictive power is by adding
some additional conditions within some model
of new physics embedding the seesaw mechanism. In this respect
quite an interesting example is represented
by the  '$SO(10)$-inspired scenario' \cite{branco},
where $SO(10)$-inspired conditions are over-imposed onto the neutrino Dirac mass matrix.
In the basis where the charged leptons mass matrix and the Majorana mass matrix are diagonal,
this is expressed in the bi-unitary parametrization as $m_D = V_L^{\dagger}\,D_{m_D}\,U_R$,
where $D_{m_D}\equiv {\rm diag}({\l_1,\l_2,\l_3})$ is the diagonalized neutrino Dirac mass matrix
and mixing angles in the unitary matrix $V_L$ are of the order
of the mixing angles in the CKM matrix.
The matrix $U_R$ can then be calculated from $V_L$, $U$ and $m_i$,
considering that, as it can be seen from the seesaw formula (\ref{seesaw}),
it provides a Takagi factorization of
$M^{-1} \equiv D^{-1}_{m_D}\,V_L\,U\,D_m\,U^T\,V_L^T\,D^{-1}_{m_D}$,
or explicitly $M^{-1} = U_R\,D_M^{-1}\,U_R^T$.
In this way the RH neutrino masses and the matrix $U_R$ are expressed in terms of the
low energy neutrino parameters, of the eigenvalues $\l_i$ and of the parameters in $V_L$.
Since one typically obtains $M_1 \sim 10^{5}\,{\rm GeV}$ and $M_{2}\sim 10^{11}\,{\rm GeV}$,
the asymmetry produced from the lightest RH neutrino decays is negligible and the
$N_2$-dominated scenario is realized \cite{SO10,SO10b}.

Imposing the leptogenesis bound
and considering that the final asymmetry does not depend on $\l_1$ and on $\l_3$, one obtains
constraints on all low energy neutrino parameters and some examples are shown in the Fig.~3
for a scan over the $2\sigma$ ranges of the allowed values of the
low energy parameters and over the parameters
in $V_L$ assumed to be $I< V_L < V_{CKM}$, where $V_{CKM}$ is the Cabibbo-Kobayashi-Maskawa
matrix \cite{SO10b}. This scenario has been also studied in a more general context
including a type II contribution to the seesaw mechanism from a triplet Higgs \cite{abada}.
\begin{figure}
\vspace*{-1mm}
\begin{center}
     \psfig{file=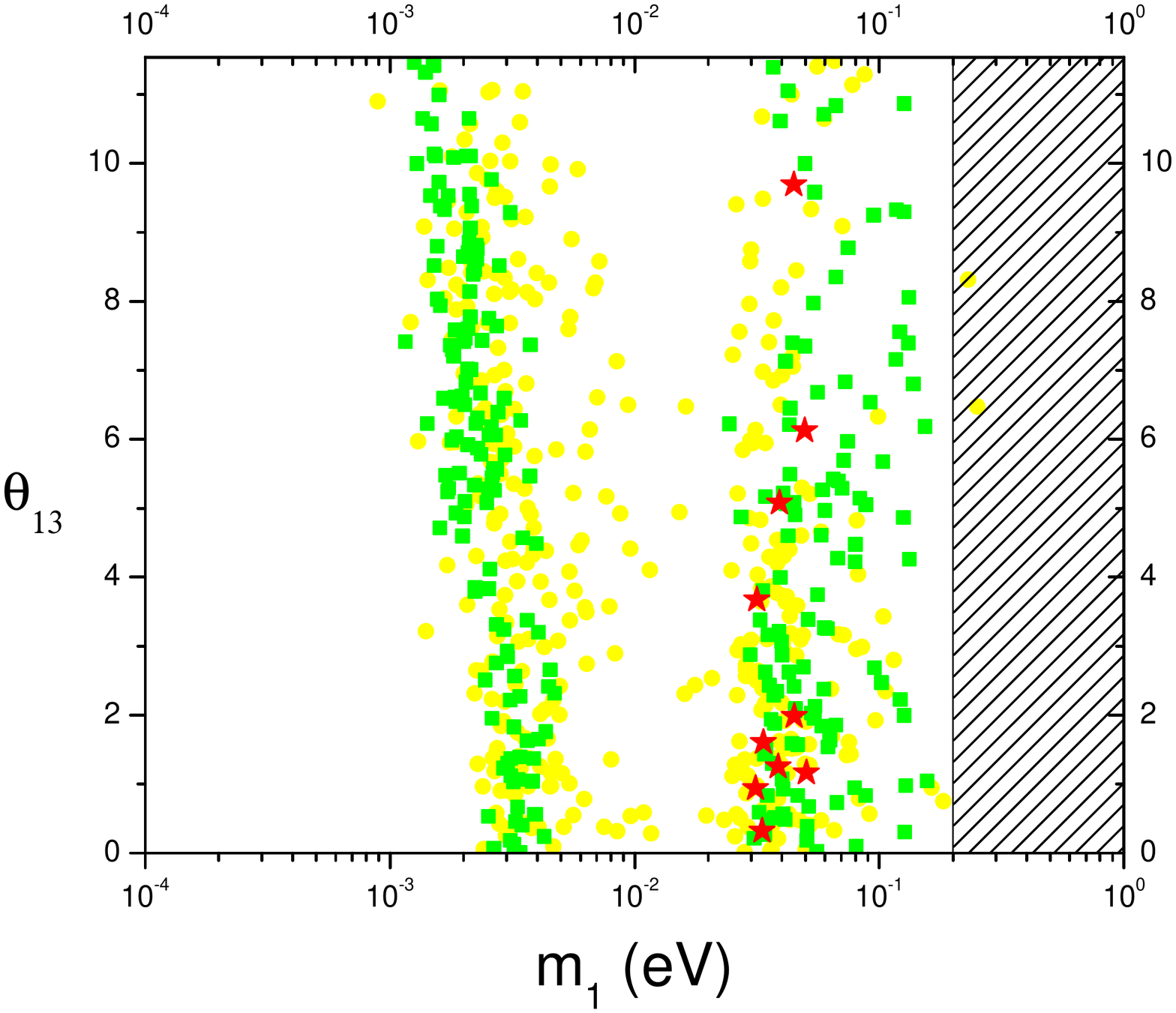,height=65mm,width=75mm} \\
     \vspace*{0mm}
      \psfig{file=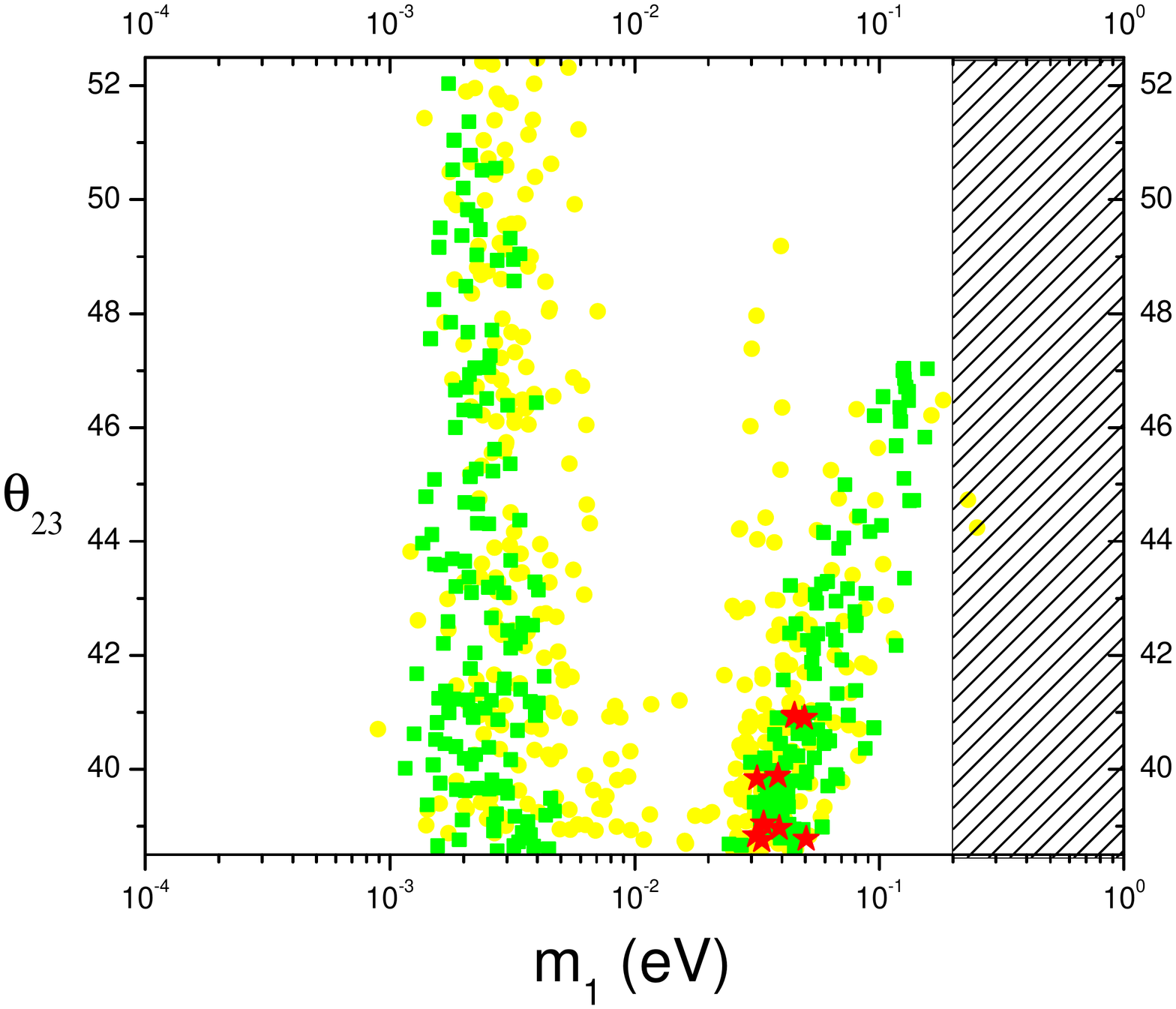,height=65mm,width=75mm} \\
      \vspace*{0mm}
       \psfig{file=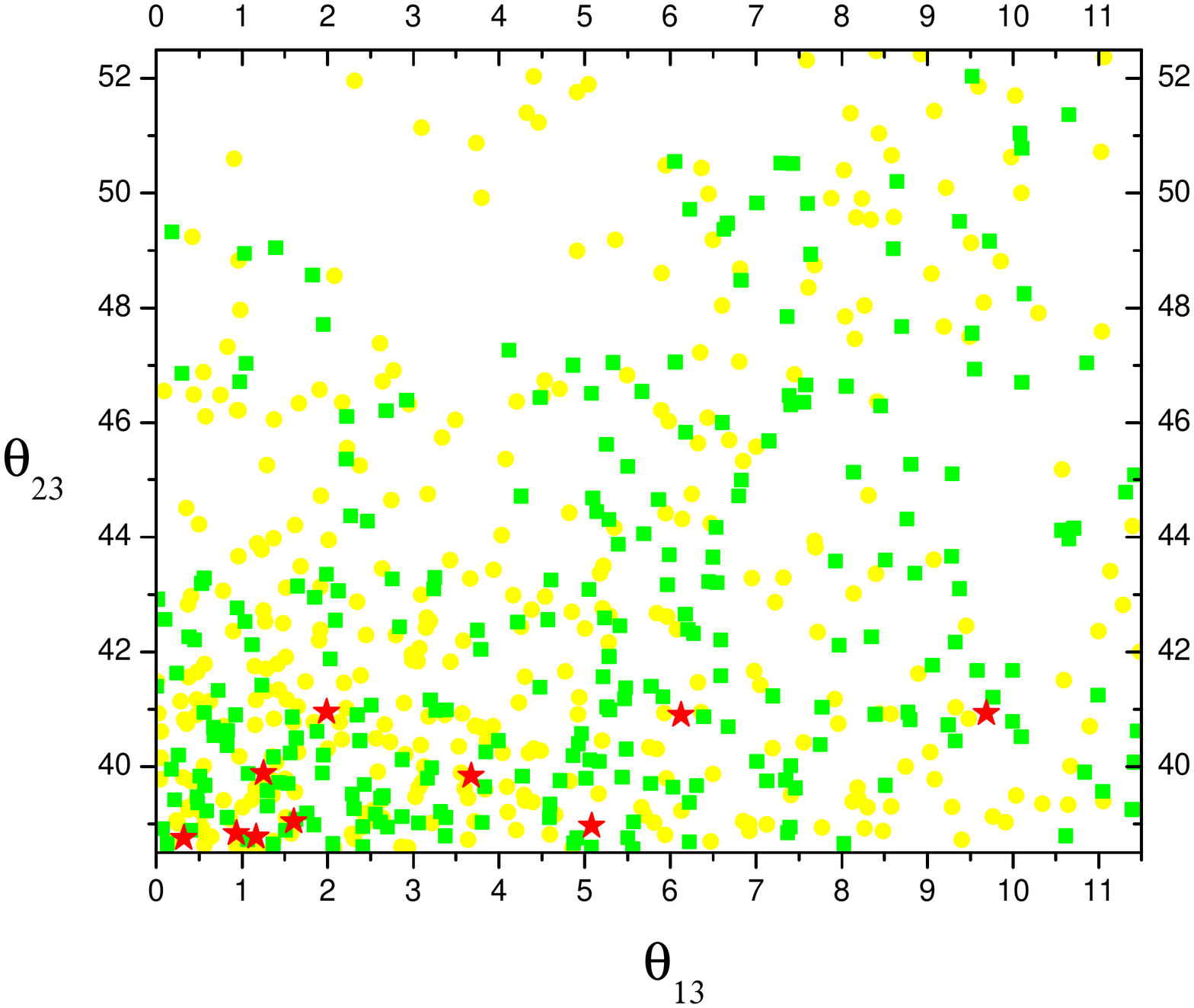,height=65mm,width=75mm}
     \vspace*{0mm}
     \caption{Constraints on some of the low energy neutrino parameters
     in the $SO(10)$-inspired scenario for normal ordering and $I< V_L < V_{CKM}$ \cite{SO10b}.}
\end{center}
\vspace*{-1mm}
\end{figure}
\subsection{Discrete flavour symmetries}

Heavy flavour effects are quite important when leptogenesis is embedded within
theories that try to explain the emerging tribimaximal mixing structure in the leptonic
mixing matrix via flavour symmetries. It has been shown in particular that
if the symmetry is unbroken then the $C\!P$ asymmetries of the RH neutrinos would exactly
vanish. On the other hand when the symmetry is broken, for the naturally expected
values of the symmetry  breaking parameters, then the observed
matter-antimatter asymmetry can be successfully reproduced \cite{manohar,feruglio}.
It is interesting that in a minimal picture based on $A4$ symmetry, one has a RH neutrino mass spectrum with
$10^{15}\,{\rm GeV} \gtrsim M_3 \gtrsim M_2 \gtrsim  M_1 \gg 10^{12}\,{\rm GeV}$. One has therefore
that all the asymmetry is produced in the unflavoured regime and that the mass spectrum
is only mildly hierarchical (it has actually the same kind of hierarchy of light neutrinos).
At the same time the small symmetry breaking imposes
a quasi-orthogonality of the three lepton quantum states produced in the RH neutrino
decays. Under these conditions the wash-out of the asymmetry produced by one RH neutrino species
from the inverse decays of a lighter RH neutrino species is essentially negligible. The final
asymmetry then receives a non negligible contribution from the decays of all three RH neutrinos species.

\subsection{Supersymmetric models}

Within a supersymmetric framework the final asymmetry within
the vanilla leptogenesis scenario undergoes  small changes \cite{proceedings}.
However, supersymmetry introduces a conceptual important issue: the stringent
lower bound on the reheat temperature, $T_{\rm RH}\gtrsim 10^{9}\,{\rm GeV}$,
is typically marginally compatible with an upper bound
from the avoidance of the gravitino problem $T_{\rm RH}\lesssim 10^{6-10}\,{\rm GeV}$, with the
exact number depending on the parameters of the model \cite{gravitino}. It is quite remarkable
that the solution of such a issue inspired an intense research activity on supersymmetric
models able to reconcile minimal leptogenesis and the gravitino problem. Of course on the
leptogenesis side, some of the discussed extensions beyond the vanilla scenario that relax the neutrino
mass bounds also relax the $T_{\rm RH}$ lower bound. However, notice that in the $N_2$ dominated
scenario, while the lower bound on $M_1$ is completely evaded, there is still a lower bound
on $T_{\rm RH}$ that is even more stringent, $T_{\rm RH}\gtrsim 6\times 10^{9}\,{\rm GeV}$ \cite{geometry}.

As we mentioned already, with flavour effects one has the possibility to relax the lower bound
on $T_{\rm RH}$ if a mild hierarchy in the RH neutrino masses
is allowed together with a mild cancelation in the seesaw formula \cite{bounds}.
However for most models, such as sequential dominated models \cite{sequential},
this solution does not work. A major modification introduced by supersymmetry
is that the critical value of the mass of the decaying RH neutrinos
setting the transition from an unflavoured regime to a two-flavour regime
and from a two-flavour regime to a three flavour regime is enhanced by a factor $\tan^2\beta$ \cite{antusch}.
This has a practical relevance in the calculation of the asymmetry within supersymmetric models
and it is quite interesting that leptogenesis becomes sensitive to such a relevant
supersymmetric parameter. Recently, a detailed analysis
mainly discussing how asymmetry is distributed among all particle species,
has shown different subtle effects in the calculation of the final asymmetry
within supersymmetric models but it just found ${\cal O}(1)$ corrections
to the final asymmetry \cite{superequilibration}.

\section{Future prospects}

In recent years, there have been important developments in leptogenesis
first of all involving a full account of (light and heavy) flavour effects
and also a deeper kinetic description accounting for quantum kinetic effects.
Many efforts are currently devoted to explore possible ways to test the seesaw mechanism
and leptogenesis. The possibility to have
models with a seesaw scale down to the TeV scale, are gaining a lot of attention,
especially in the light of the LHC and with the prospect of solving the hierarchy problem
\cite{mohapatratalk,seesawLHC}.
This possibility seems necessarily to involve non minimal leptogenesis models based on a
seesaw mechanism beyond the minimal type I \cite{petcov}.

Even within traditional high energy scale leptogenesis,  flavour effects have
opened new opportunities, or re-opened old ones, to test leptogenesis.
In a minimal leptogenesis scenario, among the many possible mass patterns, a genuine
$N_2$-dominated scenario with $M_1\ll 10^{9}\,{\rm GeV}$ and $M_2\gtrsim 10^{9}\,{\rm GeV}$,
presents some attractive features: i) the presence of a double
stage, production from $N_2$ decays and wash-out from $N_1$ inverse processes,
seems to enhance the predictive power yielding constraints on the low
energy parameters; ii) it provides a solution to the problem
of the independence of the initial conditions if the final asymmetry is
tauon dominated (in this case the constraints on the low energy parameters
become even more meaningful) \cite{preexisting};
iii) it rescues the interesting class of $SO(10)$-inspired
models leading to testable constraints on the low energy neutrino parameters.

We can fairly conclude saying that leptogenesis is experiencing a mature stage with
various interesting ideas about the possibility to test it.
Low and high energy scale models lead to
quite different phenomenological scenarios. In the first case they necessarily predict
some novel phenomenology. In the case of more conventional
high energy scale models, the naturally  expected experimental progress
in low energy neutrino experiments could uncover some
non trivial correlations among parameters. These correlations would be a trace of the
dynamical processes that led to the generation of observed matter-antimatter asymmetry
during a very early stage in the Universe history and would specifically depend on the model
of new physics embedding the seesaw mechanism.

\subsection*{Acknowledgments}

I wish to thank S.~Antusch, E.~Bertuzzo, S.~Blanchet, W.~Buchmuller, F.~Feruglio, D.~Jones,
S.~King, L.~Marzola, M.~Plumacher, G.~Raffelt, A.~Riotto for a fruitful collaboration
on leptogenesis.  I acknowledge financial support from the NExT Institute and SEPnet.

\bibliographystyle{elsarticle-num}



\end{document}